# Alpha Particle–Induced Collision Cascade Fusion

Sandeep Puri [a,c]*, Noah D'Amico [a,c], Andrew Gillespie [a], Ian Jones [a], Cuikun Lin [a], Bo Zhao[b], R. V. Duncan [a,c]

[a] *Center for Emerging Energy Sciences, Department of Physics and Astronomy, Texas Tech University, Lubbock, Texas, USA*
[b] *College of Arts and Sciences Microscopy, Texas Tech University, Lubbock, Texas, USA*
[c]*BlankSlate Innovation, LLC, Lubbock, Texas 79416*
*** Corresponding Authors:** sandpuri@ttu.edu

## Abstract

We report experimental and computational investigations of a collision-cascade mechanism to induce deuterium-deuterium (D–D) fusion. Evidence of neutron production was observed from a pressurized deuterium target exposed to energetic alpha particles emitted by a $^{210}$Po source. A 5-mCi $^{210}$Po alpha source was placed within a chamber containing pressurized deuterium gas, and neutron emission was monitored for 18 h using two Mirion SN-S $^{3}$He neutron detectors. Alpha particles incident on pressurized deuterium gas produced an average excess of ~74 neutrons after background subtraction, corresponding to a fusion neutron rate of ~2.24 n/s. with LiD in the pressurized deuterium, the average excess increased to 268 neutrons, corresponding to 8.1 n/s. Since D–D fusion branches into two equally likely pathways, these correspond to a fusion rate near 4.5 and 16.2 fusions per second, respectively. MCNP simulations incorporating experimental geometry, source activity, and detector configuration predicted neutron yields within 5.4% of the measured values and reproduced the detector response within experimental uncertainty. The tight agreement between measured and simulated neutron counts suggests that energetic alpha-particle interactions within the deuterium may contribute to measurable D–D fusion reactions through the D(d,n)$^{3}$He channel.

## 1. Introduction

Nuclear fusion has long been pursued as a promising source of clean and sustainable energy because of its high energy density and negligible greenhouse gas emissions. Over the past several decades, considerable progress has been achieved in both magnetic confinement and inertial confinement fusion.[1,2] Recent developments in high-temperature superconducting magnets have enabled the design of more compact magnetic confinement devices, while experiments at the National Ignition Facility demonstrated, for the first time, fusion target gain exceeding the energy delivered directly to the fuel capsule by the driver, marking a significant milestone in inertial confinement fusion research.[3,4] These advances have renewed interest in alternative approaches capable of enhancing fusion reaction rates under laboratory conditions.

In parallel with conventional plasma-based fusion concepts, considerable effort has been devoted to understanding mechanisms that enhance low-energy nuclear reactions through electron screening.[5,6] Measurements of deuteron-induced fusion in gaseous, metallic, and insulating targets have consistently demonstrated that the effective electron screening potential depends strongly on the host material.[7–19] In metallic environments, screening potentials substantially larger than those predicted by simple adiabatic models have been reported, leading to measurable enhancement of D–D fusion cross sections at low projectile energies. The underlying mechanisms remain an active area of research, with plasma and condensed-matter screening effects continuing to be investigated experimentally and theoretically.

Recently, collision-induced fusion mechanisms in condensed matter have attracted increasing attention. Steinetz *et al.* reported evidence of neutron production from highly deuterated titanium and erbium exposed to intense bremsstrahlung radiation, with measured neutron energies consistent with the $D(d,n)^3He$ reaction.[20] Subsequent theoretical analysis by Pines *et al.* suggested that energetic photoneutrons generated through photon-induced deuteron disintegration could elastically scatter from neighboring deuterons, producing recoil deuterons with kinetic energies on the order of tens of kiloelectronvolts.[21] These energetic recoil deuterons were proposed to undergo subsequent collisions with stationary deuterons within the lattice, thereby initiating collision-induced D–D fusion. Although the predicted fusion probability for an individual recoil deuteron remains small, the mechanism demonstrated that secondary particle collisions may provide an alternative pathway for producing fusion reactions without conventional charged-particle acceleration.

Motivated by these developments, our research group recently investigated collision-induced nuclear processes in deuterated solid-state materials subjected to neutron irradiation. The findings from this work were published in Nuclear Engineering and Technology.[22] In these studies, we investigated methods for generating energetic tritium within the proposed reaction cycle, as well as the potential for collisional acceleration of deuterium by energetic neutrons to energies sufficient for initiating fusion reactions with stationary deuterium atoms elsewhere in the lattice. It is important to emphasize that experiments involving thermal neutrons naturally introduce well-established pathways for tritium production. Any reactions that produce $^3$He may subsequently enable tritium generation via the $^3$He(n,p)T reaction. Although its cross section is much smaller, radiative capture through D(n,γ)T can also contribute to tritium formation. In the present study we investigate a collision-cascade mechanism for inducing D–D fusion in a compressed deuterium gas environment. Energetic alpha particles emitted from a $^{210}$Po source deposit energy through successive interactions within pressurized $D_2$ gas. These interactions generate recoil deuterons and initiate a cascade of secondary collisions. The resulting non-equilibrium population of energetic deuterons increases the probability of D–D encounters at energies sufficient for fusion through the D(d,p)T and D(d,n)$^3$He reaction channels. Neutron emission from the latter branch is monitored using two independent $^3$He neutron detectors.

The results obtained in the previously published study on collisional fusion are both promising and exciting. In that study, samples irradiated in the reactor produced between $5\times10^{12}$ and $6\times10^{12}$ tritium nuclei, consistent with simulations based on known physical processes. Additional fusion events may have occurred; however, the resulting byproducts were below the detection limits for tritium. The high thermal neutron flux in the reactor environment strongly supported tritium generation through the D(n,γ)T pathway. In contrast, samples irradiated in the cyclotron vault exhibited tritium production rates 2.9–5.1 times higher than those predicted by simulations, indicating promising and potentially novel reaction pathways-collisional induced fusion- worth further investigation.

The principle of collision-induced fusion is based on the generation of energetic charged particles that initiate a cascade of ionization and secondary collisions, ultimately leading to fusion events. As illustrated in **Figure 1**, high-energy alpha particles propagating through a compressed gaseous medium can produce extensive ionization tracks. These ionization cascades generate energetic secondary ions, which in turn undergo collisions with surrounding nuclei, creating localized conditions favorable for fusion reactions.

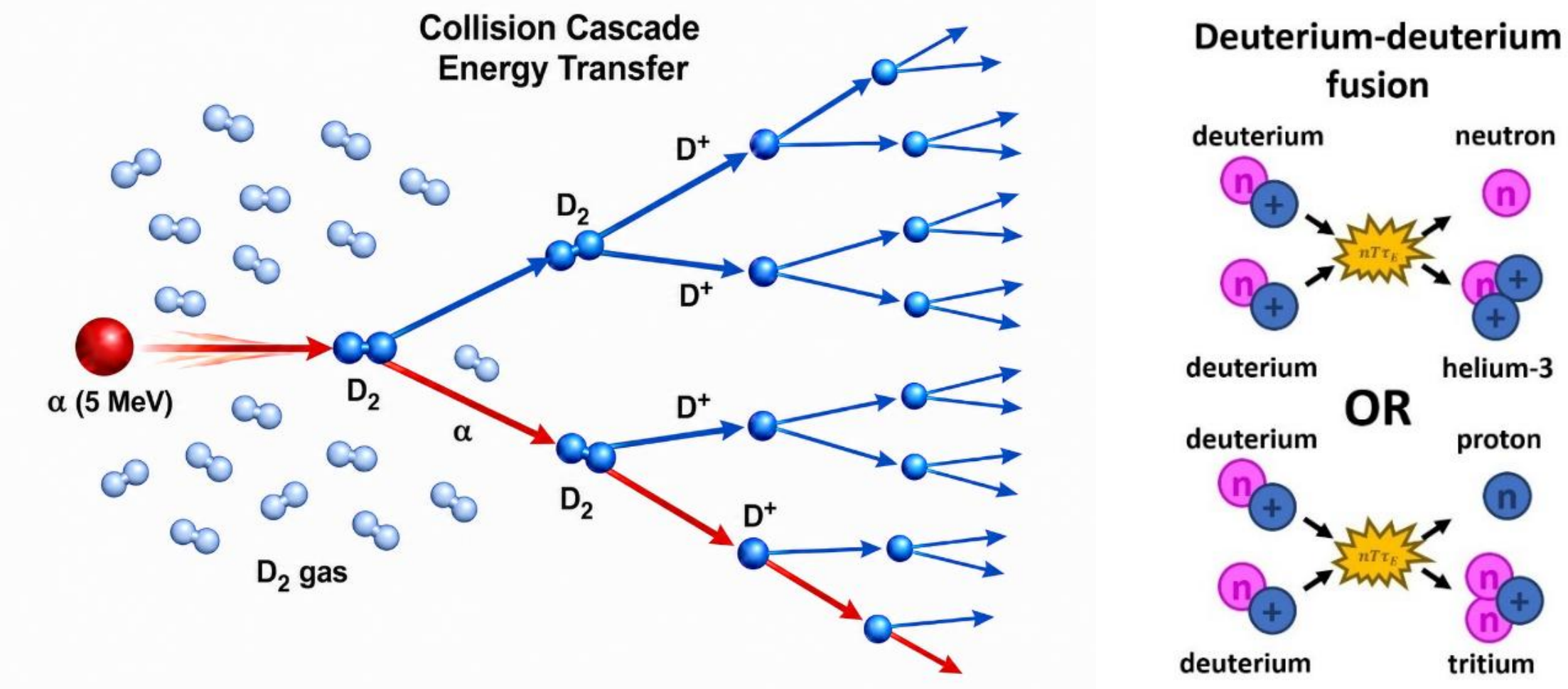

**Figure 1:** *Left*: Alpha particles incident on deuterium gas, creating an ionization cascade. *Right*: Depiction of two primary D–D fusion channels.

The resulting fusion rate depends on several key parameters, including the initial energy of the charged particles, relevant nuclear and atomic collision cross sections, and the stopping range of the particles in the target medium. For example, a 5 MeV alpha particle has a stopping range of approximately 1.7 cm in deuterium ($D_2$) gas at a pressure of 10 bar. By increasing the pressure—and thus the density—of the compressed $D_2$ gas, the stopping range is reduced, effectively confining the ionization and collision cascade within a very small spatial volume. This confinement significantly increases the local collision frequency and enhances the probability of collisionally-induced fusion. This approach enables fusion reactions to be localized within a finely confined spherical region, allowing flexibility in fuel selection. Gas mixtures such as $D_2/T_2$ or $D_2/^3He$ can be employed depending on the desired reaction channels and byproducts.

It is important to emphasize that this concept is fundamentally different from conventional neutron generators and plasma-based fusion systems. In accelerator-type neutron generators, deuterium gas is first ionized using radiofrequency excitation and then accelerated to energies on the order of 100–120 keV toward a solid metal deuteride or tritide target. These systems operate at extremely low pressures, typically in the microtorr range. Under such high-vacuum conditions, the mean free path of $D^+$ ions extend to kilometers, making collisionally-induced fusion in the gas phase effectively impossible. In contrast, the approach used in this study intentionally operates at elevated gas pressures to maximize collisional interactions and energy deposition within a confined region, enabling a fundamentally different fusion regime.

## 2. Monte Carlo Simulations for Fusion Rates in Simplified Experiment

A small prototype was developed to test D–D fusion induced by collisions with energetic alpha particles. The Monte Carlo n-Particle (MCNP)[23,24] transport simulations involve a cylindrical steel chamber (12 cm length, 3 cm diameter) filled with $D_2$ gas between 4-10 bar pressure. A source zone of ~5.5-MeV alpha particles was placed inside one end of the cylinder. A simple depiction is shown in **Figure 2**. One control simulation was performed with $H_2$ gas to test the significance of any (α,n) reactions that may have occurred within the stainless steel.

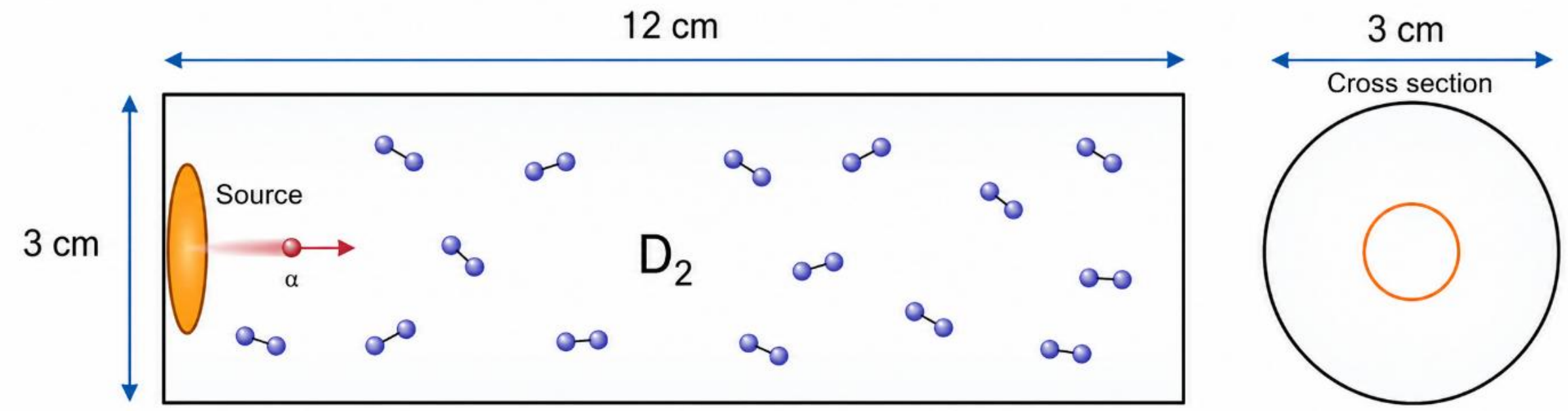

**Figure 2:** Schematic of a gas chamber and alpha particle source.

The first set of simulations set out to obtain the energy distribution for $D^+$, $T^+$, $^3He^{2+}$ ions after collisions with alpha particles and to determine the number of these ions created with respect to the input number of alpha particles. This analysis was repeated for a 100% $D_2$ loading gas using NIST densities at T = 20 °C.

On time scales near ~100 ms, all of the ions slow down and return to near thermal energies. But on the scale of microseconds, we assume collisions will deposit energies on the order of 0.1 to 4 MeV, before secondary collisions spread the energy out across other D ions. The energy spectra of relevant particle types were recorded for 1-cm tall cylindrical slices along the axis of the sample chamber.

**Table 1:** Particle creation rates with respect to the number of input alpha particles from MCNP simulations. $10^{11}$ source alpha particles of ~5.5 MeV assumed for each simulation.

| Gas and Pressure | Approx. D2 Gas Density [g/L] | *n* created | $D^+$ ions created | $T^+$ ions created | $^3He^{2+}$ ions created |
|---|---|---|---|---|---|
| 100% $D_2$, P = 4 bar | 0.6595 | 873 | 32,288,363 | 794 | 873 |
| 100% $D_2$, P = 6.5 bar | 1.0702 | 1,277 | 38,449,160 | 1,121 | 1,309 |
| 100% $D_2$, P = 10 bar | 1.6431 | 1,623 | 45,605,419 | 1,442 | 1,723 |

Note that the way the term "created" is used within MCNP pertains to the outcome of any collision. Here, many of the particles that are created are just receiving kinetic energy during the collisions with source and secondary particles. These results may be scaled to the source strength to predict expected fusion rates. We can repeat the simulations with a detector or spectrometer zone to determine the expected signal to noise ratio from an experiment.

The simulations involving alpha particles incident upon 100% $D_2$ gas used $10^{11}$ particle histories to decrease the statistical uncertainties in neutron tallies and particle creation rates. There was an additional goal of determining whether there was any directional bias for the fusion neutrons. There was a bias for the fusion neutrons to travel in the same direction as the source particles rather than being emitted in a purely isotropic manner.

The F4 tally in MCNP records the "average track length of the particle flux." This calculates the track length estimate of cell flux in units of [1/cm$^2$] and is calculated by multiplying the importance weight by the track length and is normalized by the cell volume. When multiplied by the source strength in particles per second, this results in the average flux across the tally zone in units of [particles/s/cm$^2$]. The F4 tally can be useful in estimating the particle yield reaching a volume of interest. The F4 tally may also be split into energy bins to obtain an energy spectrum of particles entering the zone. MCNP is also able to output particle tracks entering each defined volume cell.

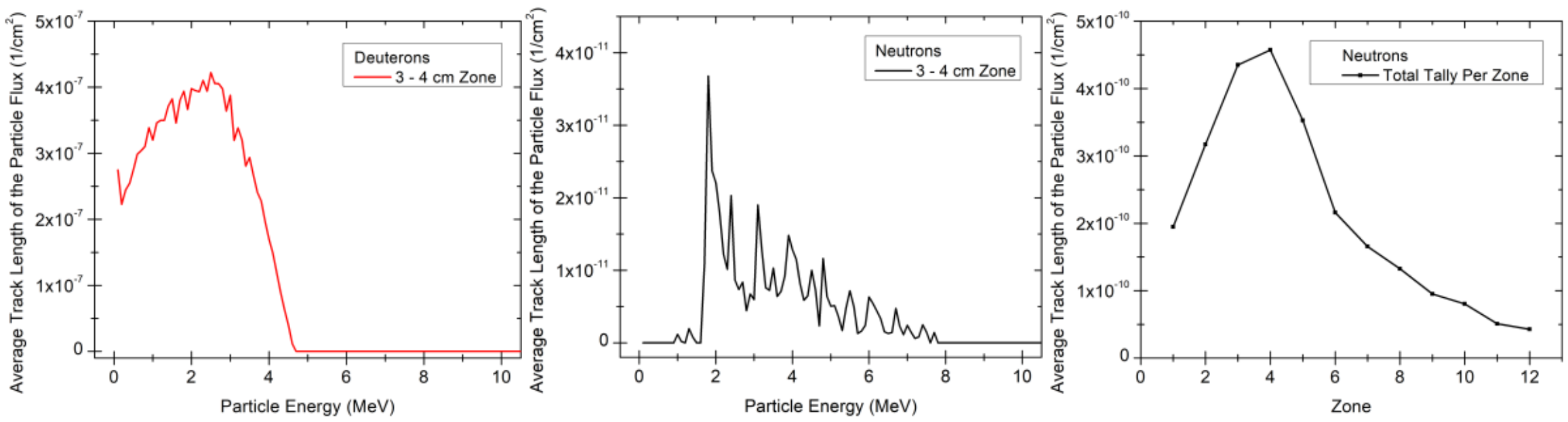

**Figure 3:** Energy distributions of n, and $D^+$ ions within the 1-cm cylindrical slices after being irradiated with the 5.5-MeV alpha particle source. *Left*: Energy spectrum for $D^+$ ions. *Middle*: Energy spectrum for neutrons. *Right*: Total neutron tally as a function of zone number (distance from source).

Alpha particles collide with deuterons that accelerate to energies up to ~5 MeV. These collide to produce fusion neutrons with reactions occurring most frequently in the region between 3 and 4 cm from the alpha source. **Figure 3** shows the energy distributions of each particle type after collisions resulting from alpha particle bombardment. Zones at distances between 2 – 5 cm display the highest rate of elastic energy transfer to deuterons. Representative spectra of the deuteron and neutron energies are presented from the 3 – 4 cm zone. For simulations using gas pressures of 4 bar, about 8.7 neutrons were created per billion source alpha particles. For simulations using gas pressures of 6.5 bar, about 12.8 neutrons were created per billion source alpha particles. For simulations using gas pressures of 10 bar, about 16.2 neutrons were created per billion source alpha particles.

A 5-mCi source emits about $1.85\times10^8$ alpha particles per second, corresponding to about $1.1988\times10^{13}$ alpha particles during the 18-hour exposure. For loading gas pressures between 4 – 10 bar, between 104,655 – 194,565 fusion neutrons were expected to be created, or between 1.6 – 3.0 neutrons per second above background levels.

Additional simplified simulations were performed to estimate the solid angle subtended by each neutron detector in comparison to the active region of the sample cell. The cylindrical detector zones were each 34 cm long with a diameter of 8.5 cm. One detector was oriented vertically, parallel with the axis of the sample cell cylinder. The second detector was oriented horizontally, perpendicular to the axis of the sample cell cylinder. Respectively, these detectors are shown on the left and right side of the sample cell in **Figure 4**. A cylindrical volume source of neutrons was used, emitted from the interior of the entire sample cell. The orientation differences and geometric constraints on their distance from the sample resulted in between 23.62 – 27.32% of the source neutrons reaching each detector. An average solid angle near 3.2 sr, or about 25.47% of the total neutron generation was used in estimating the experimental detection rate.

## 3. Experimental Evidence of Alpha Particle-Driven Collisional Fusion in $D_2$/LiD

Preliminary data was gathered using the same experimental parameters as described in **Figure 2** with the experimental setup shown in **Figure 4**. A 5-mCi alpha particle source of $^{210}$Po was placed in chamber of 6.5 bar $D_2$ gas. Two helium-3 neutron detectors (Mirion SN-S) were placed alongside the chamber with one oriented parallel to the axis of the cylindrical sample cell and the other oriented perpendicularly.

A control experiment was performed with no hydrogen cell nor $^{210}$Po. This resulted in 185-200 neutrons detected from background measurements. Other preliminary experiments were performed with different gas pressures and target materials. Approximately 193 neutrons were detected from background exposure over the course of an 18-hour measurement. The hydrogen and air control cells registered roughly 373 counts over an 18 h run — well above the source-free background. The appropriate background level for the deuterium gas measurements is therefore this source-plus-non-deuterium level of 373 counts. During the active ($D_2$ gas) experiment, 408 and 485 neutrons were detected from the chamber (~447 average). This represents a signal of roughly 74 neutrons above background.

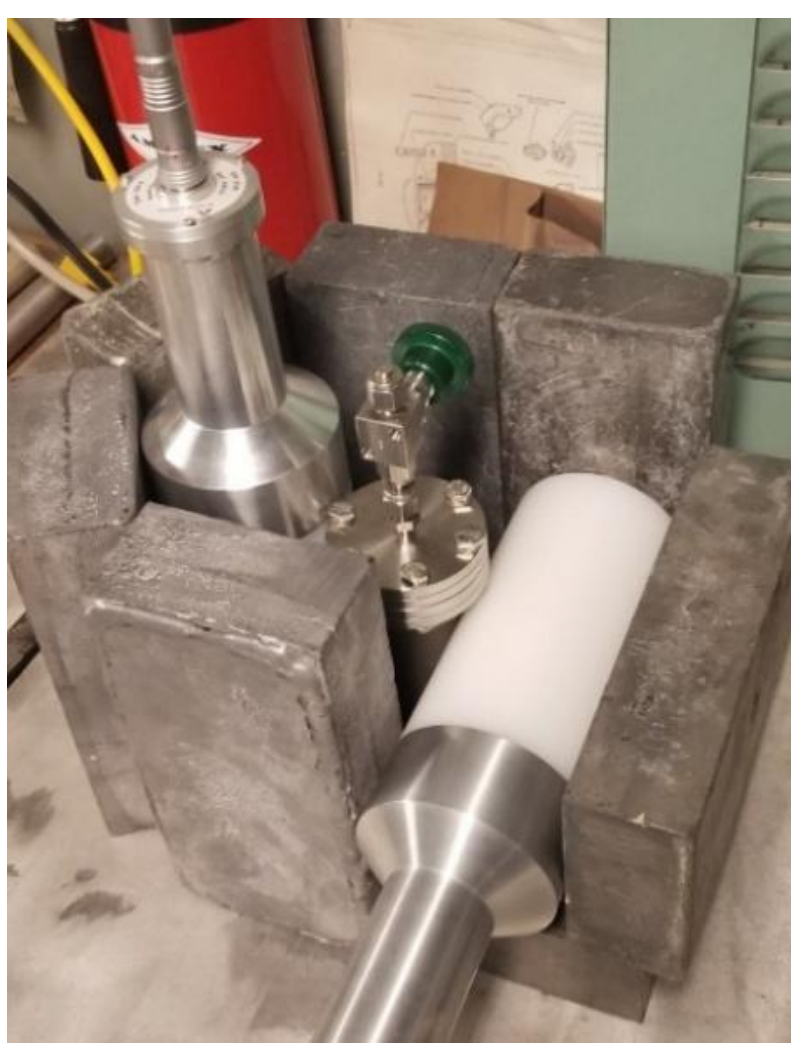

**Figure 4:** Stainless steel active experimental chamber with two neutron detectors surrounded by lead reflectors.

In experiments involving LiD and Pd in deuterium gas, a plastic cup was added to encompass the $^{210}$Po source and the sample metal (LiD or Pd). Due to limited time with the $^{210}$Po source, we were unable to run true 1:1 control experiments for the LiD and Pd samples, in which we would have filled the chamber with $H_2$ gas, and LiH would have been substituted for LiD. The LiD exceeded the background regardless of the choice of 193 or 373 counts as the true background signal. In contrast, the neutron count for the Pd sample is within the background envelope. The rapid decay and depletion of the $^{210}$Po source (half-life near 138 days) precludes the acquisition of additional background measurements. Therefore, a definitive experimental conclusion cannot be drawn regarding fusion events in the Pd sample (as shown by the 334-count average in **Table 2**). Conversely the deuterium-bearing targets ($D_2$ and LiD + $D_2$) consistently exceed the background control measurements. Results from control groups and active experiments are outlined in **Table 2**.

**Table 2**: Neutron detection from various target materials after irradiation with high-energy alpha particles.

| | Exposure Time [h] | Neutrons Detected [Counts] | | | Corresponding Neutron Rate [n/s] |
|---|---|---|---|---|---|
| | | Horiz. | Vert. | Avg. | |
| Background | 18 | 200 | 185 | 193 | - |
| $D_2$, 6.5 bar | 18 | 485 | 408 | 447 | 2.2 |
| $D_2$, 9.9 bar | 18 | 483 | 437 | 460 | 2.6 |
| $H_2$, 10 bar | 18 | 389 | 342 | 366 | - |
| Air, 0.9 bar | 18 | 400 | 361 | 381 | - |
| LiD, $D_2$, 10.8 bar | 18 | 478 | 444 | 461 | 8.1 |
| LiD, $D_2$, 10.8 bar | 109 | 2769 | 2468 | 2619 | 7.2 |
| Pd, $D_2$, 6.5 bar | 18 | 353 | 314 | 334 | 4.3 |

The final column in **Table 2** was calculated after background subtraction and based on the energy of the neutrons, cross section with the detector, detector efficiency, and solid angle subtended by each detector. The appropriate background was subtracted for each experiment type, and the remaining signal was scaled up based on a solid angle fraction of 25.5%. The SN-S detectors have an efficiency of approximately 0.3% for a $^{252}$Cf source at a distance of 1 m. Assuming an average D–D fusion neutron energy of 2.5 MeV, the $^{3}$He detection cross section for D–D fusion neutrons is approximately 2/3 of that for the $^{252}$Cf neutron spectrum, resulting in an estimated detection efficiency of ~0.2%. Thus, approximately $74/(0.255 \times 0.002) \approx 1.45 \times 10^5$ neutrons were generated in the sample after background subtraction, corresponding to a neutron production rate of ~2.24 neutrons/s.

Simulations estimated that 153,087 total neutrons were expected in this time, or ~2.36 neutrons/s. This has a percent difference of less than 5.4% compared with what was detected experimentally, which is below the 7.26% uncertainty introduced by the solid angle estimate. This difference could be due to neutrons being reflected off the lead shielding (that was not included in simulations), thus increasing the reaction rate. Further testing of the $^{3}$He detectors is necessary to confirm a more accurate estimate of their detection efficiencies.

Initial experiments using a $^{210}$Po source with LiD/$D_2$ targets have shown neutron signals significantly above background levels in our bunker facilities, and a preliminary calculation of these amplified fusion rates corresponded to around 7 – 8 n/s, comparable to those observed with the $D_2$ target but higher than control experiments using $H_2$ and air. These results suggest successful collisional fusion in the $D_2$ and LiD/$D_2$ systems, with the observed collisional ion amplification representing an important contribution of this design. The order-of-magnitude alignment between experiments and idealized simulation results offers confirmation that a collision-cascade mechanism is active and this study offers insight into the underlying mechanisms. These findings demonstrate a statistically significant neutron enhancement and motivate further investigation of collision-induced fusion processes in compressed fusion fuel (D-D, D-T, D-$^{3}$He3) environments.

Future studies will further investigate methods to enhance the collisional-fusion rate, including optimization of source activity, target pressure, exposure duration, and surface-plasma-enhanced electron-screening techniques. We will also examine D/$^{3}$He collision-induced fusion, which has a substantially larger peak fusion cross section than D–D fusion but presents additional experimental challenges because it is largely aneutronic, requiring alternative diagnostics such as charged-particle or tritium detection. In addition, MCNP simulations will be used to explore the feasibility of fission-fragment–induced collisional fusion, where ~200 MeV fission fragments may generate further collisional ion amplification and potentially increase fusion yield.

**4. Conclusion**

A pressurized deuterium gas target irradiated by a 5-mCi $^{210}$Po alpha source produced neutron counts significantly above background as measured by two independent $^{3}$He detectors. After background subtraction, approximately 145,100 excess neutrons were observed during an 18-hour exposure. The excess is statistically significant and warrants further investigation.

This study demonstrated alpha-particle-driven collisional fusion within a compressed $D_2$ gas environment. The experimental detection of a statistically significant neutron flux above background levels directly confirms the viability of the localized collision-cascade mechanism. For experiments using compressed $D_2$ gas, the empirical yield (~2.24 n/s) achieved quantitative agreement with the predictive MCNP transport simulation (2.36 n/s) within 5.4%. While the measurements are consistent with neutron-producing processes occurring within the deuterium target, additional control experiments will be necessary for future optimization. This study paves the way for future research in more advanced, higher-yield regimes, including surface-plasma-enhanced screening and exploring the feasibility of fission-fragment-driven cascades.

**5. Conflict of Interest**

The authors declare that the research was conducted in the absence of any commercial or financial relationships that could be construed as a potential conflict of interest.

**6. Author Contributions**

All authors contributed to the conception and design of the research reported here. SP wrote the first draft of the manuscript. All authors contributed to manuscript revision, read, and approved the submitted version.

**7.0 Funding**

This work was supported by the Department of Energy award No. DE-AR0001736, and Texas Tech University.

## 8. References

1. MIT-designed project achieves major advance toward fusion energy. *MIT News | Massachusetts Institute of Technology* (2021) Available at: https://news.mit.edu/2021/MIT-CFS-major-advance-toward-fusion-energy-0908 [Accessed June 10, 2026]

2. Tests show high-temperature superconducting magnets are  ready for fusion. *MIT News* (2024) Available at: https://news.mit.edu/2024/tests-show-high-temperature-superconducting-magnets-fusion-ready-0304

3. Tollefson J. US nuclear-fusion lab enters new era: achieving 'ignition' over and over. *Nature* (2024) **625**:11–12. doi:10.1038/d41586-023-04045-8

4. Tollefson J, Gibney E. Nuclear-fusion lab achieves 'ignition': what does it mean? *Nature* (2022) **612**:597–598. doi:10.1038/d41586-022-04440-7

5. Greife U, Gorris F, Junker M, Rolfs C, Zahnow D. Oppenheimer-Phillips effect and electron screening ind+ d fusion reactions. *Z Physik A - Hadrons and Nuclei* (1995) **351**:107–112. doi:10.1007/BF01292792

6. Strieder F, Rolfs C, Spitaleri C, Corvisiero P. Electron-screening effects on fusion reactions. *Naturwissenschaften* (2001) **88**:461–467. doi:10.1007/s001140100267

7. Berlinguette CP, Chiang Y-M, Munday JN, Schenkel T, Fork DK, Koningstein R, Trevithick MD. Revisiting the cold case of cold fusion. *Nature* (2019) **570**:45–51. doi:10.1038/s41586-019-1256-6

8. Raiola F, Burchard B, Fülöp Zs, Gyürky Gy, Zeng S, Cruz J, Di Leva A, Limata B, Fonseca M, Luis H, et al. "Enhanced d(d,p)t fusion reaction in metals," in *The 2nd International Conference on Nuclear Physics in Astrophysics*, eds. Z. Fülöp, G. Gyürky,  E. Somorjai (Berlin, Heidelberg: Springer Berlin Heidelberg), 79–82. doi:10.1007/3-540-32843-2_11

9. Iwamura Y, Ito T, Kasagi J, Murakami S, Saito M. Excess Energy Generation using a Nano-sized Multilayer Metal Composite and Hydrogen Gas. *Journal of Condensed Matter Nuclear Science* (2020) **33**: doi:10.70923/001c.72546

10. Iwamura Y, Itoh T, Kasagi J, Kitamura A, Takahashi A, Takahashi K, Seto R, Hatano T, Hioki T, Motohiro T, et al. Anomalous Heat Effects Induced by Metal Nano-composites and Hydrogen Gas. *Journal of Condensed Matter Nuclear Science* (2019) **29**: doi:10.70923/001c.72496

11. Kasagi J. Low energy nuclear cross sections in metals. *Surface and Coatings Technology* (2007) **201**:8574–8578. doi:10.1016/j.surfcoat.2006.01.087

12. Czerski K. Deuteron-deuteron nuclear reactions at extremely low energies. *Phys Rev C* (2022) **106**:L011601. doi:10.1103/PhysRevC.106.L011601

13. Kasagi J, Yuki H, Baba T, Noda T. Low Energy Nuclear Fusion Reactions in Solids.

14. Schenkel T, Persaud A, Wang H, Seidl PA, MacFadyen R, Nelson C, Waldron WL, Vay J-L, Deblonde G, Wen B, et al. Investigation of light ion fusion reactions with plasma discharges. *Journal of Applied Physics* (2019) **126**:203302. doi:10.1063/1.5109445

15. Chen K-Y, Maiwald J, Schauer PA, Issinski S, Garcia FH, Oldford R, Egoriti L, Higashino S, Vakili AE, Wen Y, et al. Electrochemical loading enhances deuterium fusion rates in a metal target. *Nature* (2025) **644**:640–645. doi:10.1038/s41586-025-09042-7

16. Casey DT, Weber CR, Zylstra AB, Cerjan CJ, Hartouni E, Hohenberger M, Divol L, Dearborn DS, Kabadi N, Lahmann B, et al. Towards the first plasma-electron screening experiment. *Front Phys* (2023) **10**:1057603. doi:10.3389/fphy.2022.1057603

17. Spitaleri C, Bertulani CA, Fortunato L, Vitturi A. The electron screening puzzle and nuclear clustering. *Physics Letters B* (2016) **755**:275–278. doi:10.1016/j.physletb.2016.02.019

18. Däppen W, Mussack K. Dynamic Screening in Solar and Stellar Nuclear Reactions. *Contrib Plasma Phys* (2012) **52**:149–152. doi:10.1002/ctpp.201100099

19. Shaviv NJ, Shaviv G. Obtaining the electrostatic screening from first principles. *Nuclear Physics A* (2003) **719**:C43–C51. doi:10.1016/S0375-9474(03)00956-4

20. Steinetz BM, Benyo TL, Chait A, Hendricks RC, Forsley LP, Baramsai B, Ugorowski PB, Becks MD, Pines V, Pines M, et al. Novel nuclear reactions observed in bremsstrahlung-irradiated deuterated metals. *Phys Rev C* (2020) **101**:044610. doi:10.1103/PhysRevC.101.044610

21. Pines V, Pines M, Chait A, Steinetz BM, Forsley LP, Hendricks RC, Fralick GC, Benyo TL, Baramsai B, Ugorowski PB, et al. Nuclear fusion reactions in deuterated metals. *Phys Rev C* (2020) **101**:044609. doi:10.1103/PhysRevC.101.044609

22. Gillespie AK, Lin C, Jones I, Jeffries B, Philipps JC, Puri S, Gahl J, Brockman J, Duncan RV. Enhanced tritium production in irradiated TiD2 from collisional fusion in the solid-state. *Nuclear Engineering and Technology* (2026) **58**:104031. doi:10.1016/j.net.2025.104031

23. Kulesza J, Adams T, Armstrong J, Bolding S, Brown F, Bull J, Burke T, Clark A, Forster Iii R, Giron J, et al. MCNP® Code Version 6.3.0 Theory & User Manual. (2022). doi:10.2172/1889957

24. https://mcnp.lanl.gov/; https://silverfirsoftware.com/. Available at: https://mcnp.lanl.gov/; https://silverfirsoftware.com/